\documentclass[a4paper]{article}

\usepackage[english]{babel}
\usepackage[T1]{fontenc}

\usepackage[a4paper,top=3cm,bottom=2cm,left=3cm,right=3cm,marginparwidth=1.75cm]{geometry}

\usepackage{amsmath, amssymb}
\usepackage{graphicx}
\usepackage{url}
\usepackage{hyperref}
\usepackage{authblk} 
\usepackage{pict2e}

\makeatletter
\DeclareRobustCommand{\loplus}{\mathbin{\mathpalette\dog@lsemi{+}}}
\DeclareRobustCommand{\lotimes}{\mathbin{\mathpalette\dog@lsemi{\times}}}
\DeclareRobustCommand{\roplus}{\mathbin{\mathpalette\dog@rsemi{+}}}
\DeclareRobustCommand{\rotimes}{\mathbin{\mathpalette\dog@rsemi{\times}}}

\newcommand{\dog@rsemi}[2]{\dog@semi{#1}{#2}{-90,90}}
\newcommand{\dog@lsemi}[2]{\dog@semi{#1}{#2}{270,90}}
\newcommand{\dog@semi}[3]{%
  \begingroup
  \sbox\z@{$\m@th#1#2$}%
  \setlength{\unitlength}{\dimexpr\ht\z@+\dp\z@\relax}%
  \makebox[\wd\z@]{\raisebox{-\dp\z@}{%
    \begin{picture}(1,1)
    \linethickness{\variable@rule{#1}}
    \roundcap
    \put(0.5,0.5){\makebox(0,0){\raisebox{\dp\z@}{$\m@th#1#2$}}}
    \put(0.5,0.5){\arc[#3]{0.5}}
    \end{picture}%
  }}%
  \endgroup
}
\newcommand{\variable@rule}[1]{%
  \fontdimen8  
  \ifx#1\displaystyle\textfont3\else
    \ifx#1\textstyle\textfont3\else
      \ifx#1\scriptstyle\scriptfont3\else
        \scriptscriptfont3\relax
  \fi\fi\fi
}
\makeatother

\newcommand{\txsub}[2]{ {#1}_{ \text{\tiny{#2} } } }
\newcommand{\txsup}[2]{ {#1}^{ \text{\tiny{#2} } } }
\newcommand{\mtsub}[2]{ {#1}_{ #2 } }

\newcommand{\hR}{\hat{R}}

\newcommand{\hP}{\hat{P}}

\newcommand{\bbR}{\mathbb{R}}

        \newcommand{\calI}{\mathcal{I}}   \newcommand{\calL}{\mathcal{L}} \newcommand{\calM}{\mathcal{M}}    \newcommand{\calQ}{\mathcal{Q}}  \newcommand{\calS}{\mathcal{S}}       

\newcommand{\ttk}{\mathtt{k}}

\numberwithin{equation}{subsection}

\title{\textbf{Critical points of WAdS/CFT and higher-curvature gravity} }

\author[1]{Ger\'onimo Caselli}
\author[2]{Gaston Giribet}
\author[3]{Andr\'es Goya}
\affil[1]{Departamento de F\'{\i}sica, Universidad Nacional de Rosario. {Av. Pellegrini 250, Rosario, Santa Fe, Argentina. }}
\affil[2]{Department of Physics, New York University. {726 Broadway, New York, NY10003, USA.}}
\affil[3]{Instituto de Astronom\'{\i}a y F\'{\i}sica del Espacio. {Ciudad Universitaria, IAFE, C.C. 67, Suc. 28, 1428, Buenos Aires, Argentina.}}

\begin{document}
\maketitle

\begin{abstract}
WAdS/WCFT correspondence is an interesting realization of non-AdS holography. It relates 3-dimensional Warped-Anti-de Sitter (WAdS$_3$) spaces to a special class of 2-dimensional quantum field theory with chiral scaling symmetry that acts only on right-moving modes. The latter are often called Warped Conformal Field Theories (WCFT$_2$), and their existence makes WAdS/WCFT particularly interesting as a tool to investigate a new type of 2-dimensional conformal structure. Besides, WAdS/WCFT is interesting because it enables to apply holographic techniques to the microstate counting problem of non-AdS, non-supersymmetric black holes. Asymptotically WAdS$_3$ black holes (WBH$_3$) appear as solutions of topologically massive theories, Chern-Simons theories, and many other models. Here, we explore WBH$_3\times \Sigma_{D-3}$ solutions of $D$-dimensional higher-curvature gravity, with $\Sigma_{D-3}$ being different internal manifolds, typically given by products of deformations of hyperbolic spaces, although we also consider warped products with time-dependent deformations. These geometries are solutions of the second order higher-curvature theory at special (critical) points of the parameter space, where the theory exhibits a sort of degeneracy. We argue that the dual (W)CFT at those points is actually trivial. In many respects, these critical points of WAdS$_3 \times \Sigma_{D-3}$ vacua are the squashed/stretched analogs of the AdS$_D$ Chern-Simons point of Lovelock gravity. 
\end{abstract}

\section{Introduction}

AdS/CFT holographic correspondence \cite{Maldacena:AdS.CFT} gave rise to a revolution in high-energy physics, as it gave access to the non-perturbative regime of gauge theories and gravity. Holography opened up the possibility of addressing otherwise inaccessible problems in strongly coupled quantum field theories, in relativistic hydrodynamics, in black hole thermodynamics, in high-energy scattering amplitudes, in quantum cosmology, and in many other topics in high-energy physics as well as in other areas of physics. The indubitable capability of the holographic techniques to work out the details of strongly coupled systems led to explore similar realizations in the context of condensed matter and statistical physics \cite{HHH:0803, Hartnoll:0903}. This motivated the search for non-relativistic strongly correlated systems that could in principle allow for a holographic realization. This is how gravity duals for models with anisotropic scale invariance, both with \cite{Son:0804, Balasubramanian.McGreevy:0804} and without \cite{Kachru:0808} Galilean symmetry, were rapidly proposed; these being given by the so-called Schr\"odinger and the Lifshitz spacetimes. From a broader perspective, the search for holographic realizations beyond AdS spaces has been one of the main lines of research in theoretical high-energy physics for at least twenty years: the dS/CFT correspondence \cite{dS.CFT}, the Kerr/CFT correspondence \cite{Kerr.CFT}, the celestial holography \cite{Strominger:2017zoo} and other realizations of flat space holography \cite{Barnich:2010eb} are some examples of this. Then, the question arises as to what extent the holographic paradigm can work for non-AdS scenarios and what can we learn from such adaptations. 

One of the most interesting realizations of non-AdS holography is the so-called WAdS/WCFT correspondence, which relates 3-dimensional Warped-Anti-de Sitter (WAdS$_3$) spaces to a special class of 2D quantum field theory with chiral scaling symmetry that acts only on right-moving modes. These theories are the often-called Warped Conformal Field Theories (WCFT$_2$), and they make WAdS/WCFT particularly interesting as a tool to investigate a totally new type of 2D QFT. Besides, WAdS/CFT is interesting because it enables to apply holographic techniques to the microstate counting problem of non-AdS, non-supersymmetric black holes. The original proposal for the warped version of the correspondence \cite{WAdS.CFT} was to relate the asymptotically WAdS$_3$ spaces to a CFT$_2$. This was further studied and revisited in the literature \cite{Hofman:2011zj, WCFT, Donnay.Giribet:1504, Donnay:2015vrb}, and a refined version of it was proposed in \cite{Hofman:2011zj}, where 2D theories with chiral scaling symmetry were identified as the actual dual to gravity about WAdS$_3$, cf. \cite{Song.Strominger:1109}. The local symmetries of such 2D theories include one copy of the Virasoro (vir) algebra in a semidirect sum with a $\hat{u}(1)$ current algebra, which is exactly the asymptotic isometry algebra of WAdS$_3$ spacetimes\footnote{WAdS/WCFT correspondence, together with the properties of WAdS$_3$ spaces and of the black holes that asymptote to them, have been largely studied in the recent literature; see for instance \cite{Compere.Detournay:0701, Compere.Detournay:0808, Anninos:0809, Anninos:0905, Anninos:0906, Giribet.Leston:1006, Goya:1108, Hohm.Tonni:1001, Guica:1111, Detournay:1204, Goya:1504} and references therein and thereof.} \cite{Compere.Detournay:0701, Compere.Detournay:0808}. This symmetry algebra differs from the algebra vir $\oplus $ vir that generates the standard 2D local conformal transformations; however, as argued in \cite{WCFT}, in some respects the former is equally powerful in constraining the theory. In particular, it permits to work out a microscopic counting of black hole microstates in WAdS$_3$ spacetime by means of a Cardy type formula \cite{WAdS.CFT}. In \cite{Donnay.Giribet:1504} it was argued that, by means of a non-local transformation, the microstate counting of black holes in WAdS$_3$ space can also be organized in terms of standard CFT$_2$, yielding equivalent results.

WAdS$_3$ spacetimes are stretched or squashed deformations of AdS$_3$ spacetime which have four Killing vectors generating the isometry group $SL(2,\mathbb{R}) \otimes U(1)$. This can be regarded as a minimal symmetry breaking of the AdS$_3$ isometry group $SO(2,2)\simeq SL(2,\bbR) \otimes SL(2,\bbR)$ down to $SL(2,\bbR) \otimes U(1)$. Provided suitable boundary conditions are imposed, the asymptotic isometry group of WAdS$_3$ is generated by vir $\loplus \, \hat{u}(1)$, exactly the same symmetry that appears in Kerr/CFT. This is far from being an accident since, as we will review later, a particular case of WAdS$_3$ naturally emerges in the near horizon limit of 4D extremal black holes \cite{Bengtsson.Sandin:0509}. Besides, WAdS$_3$ spaces also appear in other contexts: they are solutions of topologically massive gravity (TMG) \cite{Moussa.Clement:9602, Clement:0301, Moussa:0303, Bouchareb.Clement:0706} and of the so-called new massive gravity (NMG) \cite{Clement:0902}; they also appear in string theory \cite{Detournay:1007, Azeyanagi:2012zd}, in theories with additional massive spin-2 fields, \cite{Goya:1406}, in higher-spin theories \cite{Gary:2012ms}, in Chern-Simons (CS) theories of lower-spin \cite{Hofman.Rollier:1411, Azeyanagi:2018har}, and in even more exotic gravity models \cite{Chernicoff:2018hpb}. Here, we will see that the WAdS$_3$ spaces also appear as geometric factors of solutions of higher-dimensional, higher-curvature gravity theories at critical points.

Critical points are curves of the parameter space of a gravity theory for which the dual CFT exhibits special properties. Typically, this leads to simplifications that permit to solve some specific problem in the CFT. For example, critical points are points at which the dual CFT becomes either chiral, or factorizable, or topological, or even trivial, or at least remarkably simple and tractable in some way. Some of the properties that the holographic theories exhibit at the critical points are the vanishing of the central charge of the boundary theory, or the emergence of bulk logarithmic modes that demands strong boundary conditions to render the dual CFT unitary, or the degeneracy of the gravity vacua. A concrete example of this is the chiral point of TMG \cite{Chiral.Gravity, Chiral.Gravity.2, Chiral.Gravity.Conjecture}, at which the right central charge, $c_R$, vanishes and new solutions appear \cite{Grumiller.Johansson:0808, GAY:0811, Compere:1006}. A similar example is 3D NMG \cite{NMG, More.NMG} with a graviton mass that equals one half of the AdS$_3$ curvature, leading to a dual CFT with no diffeomorphism anomaly and no Weyl anomaly, i.e. $c=0$, cf. \cite{Grumiller.Hohm:0911, Liu.Sun:NoteNMG, Liu.Sun:0904.GCG, Goya:1108}. Other examples are the Critical Gravity in four \cite{Lu.Pope:1101} and higher \cite{Deser:1101} dimensions, for which the black hole states have vanishing conserved charges. However, the best studied example of a critical point in higher dimensions is probably the CS point of 5D Einstein-Gauss-Bonnet gravity \cite{Zanelli:2005sa}, which is special in many respects. This corresponds to the curve of the parameter space on which the Einstein-Gauss-Bonnet gravity theory exhibits a unique maximally symmetric vacuum and the action of the theory can be expressed as a 5D CS gauge theory for the group $SO(2,4)$. In the notation of \cite{Brigante:2008gz}, this corresponds to $\lambda_{\text{GB}}= 1/4$ (in our notation, this corresponds to $\alpha \Lambda=-3/4$). This is the point where the shear viscosity to entropy density ratio, $\eta / s$, in the 4-dimensional theory vanishes, as well as the central charge $c$ -- while the other 4D central charge, $a$, takes a negative value--. While the critical point $\lambda_{\text{GB}}=1/4$ lies outside the segment of the parameter space in which the gravity theory is free of causality problems, the value $\lambda_{\text{GB}}=1/4$ itself cannot be excluded by the very same perturbative arguments as the 5D CS gravity lacks of linearized local degrees of freedom around AdS$_5$. Also at this point, $\alpha \Lambda =-3/4$, the theory exhibits degeneracy around other vacua; for example, there, both Schr\"odinger and Lifshitz spaces solve the field equations for arbitrary values of the dynamical exponent $z$ \cite{ABGGH:0909}, which is a remarkable fact that seems to imply something special about the non-renormalizability of that exponent \cite{Adams:2008zk}. Here, we will observe a similar phenomenon occurring for WAdS$_3$ vacua. More precisely, we will see that the WAdS$_3\times \Sigma_{D-3}$ vacua of the quadratic Einstein-Gauss-Bonnet gravity theory, which for $D=5$ appear when $\alpha \Lambda =-1/4$, exhibit degeneracy in the parameters that control the squashing/stretched deformation of the space and its curvature radius. In this sector, the theory behaves effectively as a topological theory whose dual WCFT$_2$ turns out to be trivial.

The paper is organized as follows: In section 2, we review the geometry of WAdS$_3$ spaces and of WBH$_3$ black holes. In section 3, we construct WBH$_3\times \Sigma_2$ solutions in $D=5$ dimensions; we discuss the computation of the black hole entropy and conserved charges, which happen to be zero. In section 4 we generalize these solutions to higher dimensions, considering different types of compactifications. We conclude that all these WAdS$_3\times \Sigma_{D-3}$ vacua are dual to theories that are trivial, with vanishing Virasoro central charge and Kac-Moody level.

\section{Warped AdS spaces}

\subsection{Hyperbolic \texorpdfstring{WAdS$_3$}{WAdS3}}

As mentioned in the introduction, the WAdS$_3$ spaces are stretched or squashed deformations of AdS$_3$. This aspect of these geometries is well-understood if AdS$_3$ space is written as a Hopf fibration over AdS$_2$ (equation (\ref{TheWAdS3}) below with $\nu=1$), cf. \cite{Witten:AdS.CFT}. WAdS$_3$ appears simply as a deformation of that fibration. Actually, AdS$_3$ appears as a particular case of WAdS$_3$, the case for which the warping deformation vanishes ($\nu = 1$ in the notation used below and in \cite{WAdS.CFT}).

The WAdS$_3$ spaces are classified in three different classes, each of them exhibiting different causal properties: One of these classes is the hyperbolic WAdS$_3$, also known as spacelike WAdS$_3$. This can easily be thought of as a warped deformation of AdS$_3$ and is the one usually considered in WAdS/WCFT holography. A different class are the elliptic WAdS$_3$ spaces, or timelike WAdS$_3$ spaces, which correspond to the 3-dimensional sections of the G\"odel solution of 4-dimensional cosmological Einstein equations. These spaces present closed timelike curves, which are inherited from its 4-dimensional GR embedding. The third class is an intermediate case, called the parabolic (or null) WAdS$_3$. This is closely related to the Schr\"odinger geometries studied in the context of non-relativistic holography. All these spaces have four Killing vectors, generating a $SL(2,\mathbb{R})\otimes U(1)$ isometry group.

As we commented in the introduction, one of the contexts in which hyperbolic WAdS$_3$ spaces naturally appear is in the study of the near horizon geometry of rapidly rotating black holes \cite{Bengtsson.Sandin:0509}, cf. \cite{WAdS.CFT}. In fact, if one considers the near horizon limit of an extremal Kerr black hole as one does in Kerr/CFT \cite{Kerr.CFT}, then one finds the 4-dimensional geometry called NHEK \cite{Bardeen:1999px}; namely
\begin{equation}
ds^2_{\text{NHEK}} = \Omega^2(\theta)\, \Big( -(\rho^2+1)d\tau^2+\frac{d\rho^2}{(\rho^2 +1)}+\Upsilon^2(\theta)\, (d\varphi\, +\, \rho\, d\tau )^2+d\theta ^2\Big)\label{LaNHEK}
\end{equation}
with
\begin{equation}
\Omega^2(\theta )= J(1+\cos^2\theta)  \, , \ \ \ \  \Upsilon (\theta) = \frac{2\sin\theta}{1+\cos^2\theta}\, ,
\end{equation}
where $\tau \in \mathbb{R}$, $\rho\in \mathbb{R}_{\geq 0}$, $\varphi \in [0, 2\pi]$, and $\theta \in [0,\pi]$. Here, $\theta $ corresponds to the azimuthal angle, and $J>0$ is the absolute value of the angular momentum of the black hole, which rotates around the axis $\theta = 0\sim \pi $. This implies $J\leq \Omega^2(\theta)\leq 2J$ and $0\leq \Upsilon^2(\theta) \leq 4$ for all $\theta$. In an appropriate system of coordinates, the 3-dimensional metric of the spacelike WAdS$_3$ is given by evaluating the 4-dimensional NHEK metric (\ref{LaNHEK}) at constant $\theta = \theta_0$; namely
\begin{equation}
ds^2_{\text{WAdS}} = \ell^2 \Big( -(\rho^2+1)d\tau^2+\frac{d\rho^2}{(\rho^2 +1)}+\frac{4\nu^2}{\nu^2+3} (d\varphi+\, \rho\, d\tau )^2\Big)\, ,\label{TheWAdS3}
\end{equation}
where $\ell^2=\Omega^2(\theta_0)$, and where $\nu $ is a convenient variable to parameterize the warping factor $\Upsilon^2(\theta_0)$; as said, $\nu=1$ corresponds to the undeformed AdS$_3$ ($\Upsilon^2=1$), while $\nu >1$ corresponds to the hyperbolic WAdS$_3$ ($\Upsilon^2>1$); the case $\nu =0$ is also special, since in that case, after rescaling as $\varphi \to \varphi /\nu $, the geometry becomes locally equivalent to AdS$_2 \otimes \mathbb{R}$. All these spaces have constant curvature invariants, some of which read
\begin{equation}
R_{A}^{\ A}=-\frac{6}{\ell^2} \, , \ \ \ 
R_{A}^{\ B} R_{B}^{\ A}=\frac{6}{\ell^4} (\nu^4-2\nu^2+3) \, , \ \ \ 
R_{A}^{\ B} R_{B}^{\ C } R_{C}^{\ A }=-\frac{6}{\ell^6}  (\nu^6+3\nu^4-9\nu^2+9)\, , \, \ \  ... \nonumber
\end{equation}
Nevertheless, for $\nu^2\neq 1$ the spaces are not of constant Riemannian curvature. For generic $\nu $ the WAdS$_3$ spaces are neither conformally flat nor asymptotically locally AdS$_3$. This can easily be seen in an appropriate coordinate system. In global coordinates, spacelike WAdS$_3$ can also be written as follows
\begin{equation}
\label{WAdSsp}
ds^2_{\text{WAdS}} = dt^2 - 2\nu r \;dt d\phi
	 + \frac{3}{4} (\nu^2-1) \, r^2 \;d\phi^2 
	 + \frac{\ell^2}{(\nu^2+3) r^2} \;dr^2  \,,
\end{equation}
with $t\in \mathbb{R}$, $r\in \mathbb{R}_{\geq 0}$, $\phi \in [0,2\pi ]$. For different coordinate system, see reference \cite{WAdS.CFT}; for coordinate systems for the elliptic WAdS$_3$, see reference \cite{Goya:1504}.

\subsection{Black holes in \texorpdfstring{WAdS$_3$}{WAdS3} }

WAdS$_3$ spaces admit black hole solutions that asymptote to them, cf. \cite{Moussa.Clement:9602, Bouchareb.Clement:0706,  Clement:0301, Moussa:0303}. In a convenient coordinate system, the metric of these black holes can be written as follows
\begin{equation}
\label{WAdSbh}
\begin{split}
ds^2_{\text{WBH}}
 &= dt^2 - \left( 2\nu r - \sqrt{r_+ r_- (\nu^2+3)} \right) \;dt d\phi \:+ \frac{\ell^2}{(\nu^2+3) (r-r_+)(r-r_-)} \;dr^2 \:+\\
	& \ \ \ \  \frac{r}{4}\left( 3(\nu^2-1)r  (\nu^2+3) (r_+ + r_-) - 4\nu\sqrt{(\nu^2+3) r_+ r_- } \right) \;d\phi^2 
 \,,
\end{split}
\end{equation}
where $r_-$ and $r_+$, provided they are both positive, describe the location of the inner Killing horizon and of the outer event horizon, respectively. These correspond to integration constant of the solution. Here, $t\in \mathbb{R}$, $r\in \mathbb{R}_{\geq 0}$, $\phi \in [0,2\pi ]$.

The Hawking temperature of the WAdS$_3$ black holes (WBH$_3$) can be computed by standard techniques, yielding
\begin{equation}
    \label{WAdSbhtemp}
    \txsub{T}{H} = \frac{(\nu^2+3)}{4\pi\ell} \dfrac{\, (r_+ -r_-)}{ (2\nu r_+ - \sqrt{(\nu^2+3) r_+ r_-})} \,.
\end{equation}

As it happens with the BTZ black holes and AdS$_3$ space, the WBH$_3$ \eqref{WAdSbh} are discrete quotients of hyperbolic WAdS$_3$ space \cite{WAdS.CFT}. This orbifold construction leads to define a right-mover and left-mover temperatures as the inverse of the identification periods; these are
\begin{equation}
T_L = \frac{(\nu^2 +3 )}{8\pi \ell} (r_++r_--\nu^{-1}\sqrt{(\nu^2+3)r_+r_-})\, , \ \ \ \ 
T_R = \frac{(\nu^2 +3 )}{8\pi \ell} (r_+-r_-)\, ,
\end{equation}
respectively. Then, we have the relation \begin{equation}
\txsub{T}{H}=\frac{(\nu^2+3)}{4\pi\ell \nu}\frac{T_R}{T_R+T_L}\, .
\end{equation}
When $r_+=r_-=0$, both $T_R$ and $T_L$ vanish, and we get the {\it empty} space \eqref{WAdSsp};
namely, the hyperbolic WAdS$_3$. Besides, the black holes (\ref{WAdSbh}) are, not only locally equivalent to WAdS$_3$, but, provided suitable boundary conditions are prescribed, they are also asymptotically WAdS$_3$. Such asymptotic boundary conditions are prescribed at large $r$ and are those preserved by the following asymptotic Killing vectors 
\begin{eqnarray}
L_n  &=&  e^{-\frac{in\phi}{\ell}} \Big( \frac{2\nu \ell^2}{\nu^2+3}\, \partial_t -in\, r\, \partial_r - \ell\, \partial _{\phi} \Big)\, +\, ... \label{Ln} \\  T_n &=& e^{-\frac{in\phi}{\ell}} \, \ell \, \partial_t\, +\, ... \label{Tn}
\end{eqnarray}
where the ellipsis stand for subleading orders in $1/r$, namely $\mathcal{O}(1/r^2)\times \partial_{\phi }$
, $\mathcal{O}(1/r)\times \partial_{t}$, $\mathcal{O}(1)\times \partial_{r}$. The Killing vectors $L_0$ and $T_0$ generate the exact $U(1) \times U(1)$ isometry of the black hole background, while $L_{\pm 1}$ complete the $SL(2,\mathbb{R})$ factor of the isometry group of global WAdS$_3$. The full set of $L_n$, $T_n$ generate the infinite-dimensional algebra
\begin{equation}
\{ L_m , L_n \} = i(n-m)L_{n+m}\, , \ \ \ \{ L_m , T_n \} = in\, T_{n+m}\, , \ \ \ \{ T_m , T_n \} = 0\, , \label{Algebrita}
\end{equation}
which is the Witt algebra in semidirect sum with the loop algebra $u(1) \otimes C^{\infty }(S^1)$. Through the Sugawara construction, this algebra produces two mutually commuting copies of Witt algebra. This was exploited in \cite{Donnay.Giribet:1504} to work out the microstate counting of WBH$_3$ from the standard CFT$_2$ perspective. The Noether charges associated to the asymptotic symmetries generated by (\ref{Ln})-(\ref{Tn}) also form an algebra which, generically, turns out to be a central extension of (\ref{Algebrita}), resulting in vir $\loplus \hat{u}(1)$, i.e. a Virasoro algebra in semidirect sum with an affine Kac-Moody algebra. However, we will argue that this is not the case for the critical points we study here: At the critical points of the $D$-dimensional higher-curvature theories we will consider, at which solutions of the form WBH$_3 \times \Sigma_{D-3}$ exhibit degeneracy in the squashing/stretching parameters $\nu $, both the Virasoro central charge and the Kac-Moody level of the dual theory vanish. 

\section{WAdS vacua in higher-curvature gravity}

\subsection{Gravity action and boundary terms}

In a $D$-dimensional higher-curvature gravity theory, we will consider solutions of the form 
\begin{equation}
\mathcal{M}=\text{WBH}_3 \times \Sigma_{D-3}\, , 
\end{equation}
with WBH$_3$ being asymptotically WAdS$_3$ black holes, and $\Sigma_{D-3}$ being a negative curvature manifold consisting of a product of locally hyperbolic spaces and tori. The simplest higher-curvature model admitting such solutions is Lovelock theory of gravity, namely the most general torsion-free metric theory of gravity yielding covariantly conserved field equations of second order. This theory propagates a single massless spin-2 mode, and, in virtue of that, it is well-behaved in many aspects. In $D\leq 4$ the theory coincides with Einstein gravity (eq. (\ref{EGB.action}) below), while it includes higher-curvature terms for $D\geq 5$. For $D=5$ and $D=6$ the action of Lovelock theory reduces to the quadratic Einstein-Gauss-Bonnet gravity action, usually considered in the context of holography. For $D\geq 7$ the theory also admits terms that are cubic in the curvature, and quartic orders appear for $D>8$. Here we will restrict the analysis to the quadratic action since this case suffices to support the backgrounds we are interested in. 

As said, we will be concerned with backgrounds of the form WBH$_3 \times\Sigma_{D-3}$. The simplest cases will be given by direct products of locally WAdS$_3$ and $(D-3)$-dimensional maximally symmetric spaces of constant curvature $\mathtt{k}$. Consider the ansatz
\begin{equation}
    \label{ansatz}
    ds^2 = {g}^{\text{(WAdS)}}_{ab} dx^{a}dx^{b} + {g}^{(\Sigma)}_{ij} dx^{i}dx^{j} \,,
\end{equation}
where ${g}^{\text{(WAdS)}}_{ab}$ are the components of the metric (\ref{TheWAdS3}) and ${g}^{(\Sigma)}_{ij}$ are the components of a space of constant curvature $\mathtt{k}$; namely
\begin{equation}
    \label{Sigma}
    {g}^{(\Sigma)}_{ij} dx^{i}dx^{j} = \frac{L^2\:\delta_{ij} \, dx^i dx^j}{\left( 1+\frac{\ttk}{4}\delta_{kl}x^k x^l  \right)^2}  \,.
\end{equation}
Here, $a , b , \, ... \, =0,1,2$, while $i,j, k, l,\, ...\, = 1, 2, ... \, ,D-3$.

The action of quadratic Lovelock theory is given by the Einstein-Gauss-Bonnet action\footnote{Our conventions follows \cite{Carroll.GR}, e.g. $[ \nabla_{M}\,, \nabla_{N} ] V^{P} = R^{P}_{\: Q MN} \, V^{Q}$, $R_{MN} = R^{Q}_{\: M Q N}$, $R = R^{M}_{M}$.}
\begin{equation}
\label{EGB.action}
    {\calI}_{\mathcal{M}} = \frac{1}{16\pi G} \int_{\calM} d^Dx \sqrt{-g} \left( R - 2 \Lambda + \alpha ( R^2 -4 R_{MN}R^{MN} +R_{MNPQ}R^{MNPQ}) \right) \,,
\end{equation}
supplemented with boundary terms, see (\ref{EGB.boundaryterms}) below. Here, $M,N,P,Q,...=0,1,2,...\,, D-1$.

The corresponding field equations read
\begin{equation}
\label{EGB.eom}
	\begin{split}
	0 = G_{MN} + \Lambda g_{MN} + \alpha \Big( &2 R_{MPQS} R_N{}^{PQS} - 4 R_{MPNQ} R^{PQ} - 4 R_{MS} R_N^S + 2 R R_{MN}\\
	    &  - \frac{1}{2} ( R^2 - 4 R_{PQ}R^{PQ} + R_{PQST}R^{PQST} ) g_{MN} \Big) \,,
	\end{split}
\end{equation}
with $G_{MN}=R_{MN}-\frac 12 Rg_{MN}$.

Being a field theory of second order, the variational principle is defined from (\ref{EGB.action}) in the usual way. This requires the inclusion of a generalized Gibbons-Hawking term. In other words, we must supplement \eqref{EGB.action} with boundary terms to guarantee a well-posed variational principle subject to Dirichlet boundary conditions on $\partial\calM$, cf. \cite{Davis:EGB.JunctionConditions}. The appropriate boundary terms are
\begin{equation}\label{EGB.boundaryterms}
    \mtsub{\calI}{\partial\calM} = -\frac{1}{8\pi G} \int_{\partial\calM} d^{D-1}x \sqrt{-h} \left( K + 2 \alpha (J - 2\hat{G}^{\mu\nu} K_{\mu\nu} \right) \,,
\end{equation}
where $K_{\mu\nu}$ is the extrinsic curvature on $\partial\calM$, $h_{\mu\nu}$ is the induced metric on $\partial\calM$, hatted tensors such as $\hat{G}_{\mu\nu}$ are constructed with the induced metric $h_{\mu\nu}$, and $J$ is the trace of tensor
\begin{equation}
\label{Jtensor}
    J_{\mu\nu} = \frac{1}{3} \left( 2 K K_{\mu\rho} K^\rho_\nu + K_{\rho\sigma} K^{\rho\sigma} K_{\mu\nu} - 2 K_{\mu\rho} K^{\rho\sigma} K_{\sigma\nu} - K^2 K_{\mu\nu} \right) \,.
\end{equation}
{Coordinates on $\partial\calM$ are denoted $ x^\mu $ where $\mu = 0, 1, \ldots D-2$.}

\subsection{Field equations and \texorpdfstring{WAdS$_3$}{WAdS3} vacua}

Let us consider first the 5-dimensional case ($D=5$). This will serve as working example throughout this section of the paper. In the next section, we will see how the results also apply for $D\geq 5$. 

Replacing the ansatz \eqref{ansatz} into the field equations \eqref{EGB.eom}, the latter turn into a simple system of algebraic equations; namely
\begin{eqnarray}
    \label{eom5D}
	E_t^{\:t} &=& \frac{3-2\nu^2}{\ell^2} \left( 1+\frac{4\alpha \ttk}{L^2} \right) + \Lambda - \frac{\ttk}{L^2} = 0 \\
	E_\phi^{\:t} &=&  \frac{3\nu r}{\ell^2} (\nu^2-1) \left( 1+\frac{4\alpha \ttk}{L^2} \right) = 0 \\
	E_\phi^{\:\phi} &=& E_r^{\:r} = \frac{\nu^2}{\ell^2} \left( 1+\frac{4\alpha \ttk}{L^2} \right) + \Lambda - \frac{\ttk}{L^2} = 0\\
	E_x^{\:x} &=& E_y^{\:y} = \Lambda + \frac{3}{\ell^2} = 0 \,.
\end{eqnarray}
Thus, we need to find the appropriate choice of the parameters $ \Lambda\,, \alpha \,, L $ and $ \ttk $ that solves this system. In order to do that, we need to distinguish between two cases: Let us consider first the case $\nu^2 =1$, which corresponds to AdS$_3 \times \Sigma_{2}$ vacua. In this case, we find
\begin{equation}
	 \Lambda=-\frac{3}{\ell^2}\;,\; \;\alpha=\frac{\ell^2}{4}+\frac{L^2}{2\ttk }\;,\; \; \ttk=\pm 1\,.
\end{equation}
However, the case of our interest is actually $ \nu^2 \neq 1$, which yields
\begin{equation}
\label{5Dsol}
	 \Lambda=-\frac{3}{\ell^2}\;,\;\;\alpha=\frac{\ell^2}{12}\;,\;\; \ttk=-1\;,\;\;L^2=\frac{\ell^2}{{3}}  \,,
\end{equation}
where we see that the cosmological constant, $\Lambda $, is negative, the coupling constant of the curvature square terms, $\alpha $, is positive, and the internal manifold has negative curvature, $\ttk=-1$, and therefore we choose the quotient $\Sigma_{2}=\mathbb{H}_2/\Gamma$, with $\Gamma $ being a Fuchsian subgroup. The remarkable fact is that there is no restriction for the squashing/stretching parameter $\nu$, which here appears as a sort of zero-mode that controls the shape of the fibration in (\ref{TheWAdS3}). It is also worth mentioning that this degeneracy appears on the curve
\begin{equation}
\alpha \Lambda = -\frac{1}{4}
\end{equation}
of the parameter space, which differs by a factor $3$ from the 5D CS point.

\subsection{On-shell action for the WAdS vacua}

Now we can evaluate the 5-dimensional action on-shell for the WAdS$_3\times \, \mathbb{H}_2/\Gamma $ ansatz. Surprisingly, everything combines in a way that the different pieces of the action $\calI = {\calI}_{\mathcal{M}} + \mtsub{\calI}{\partial\calM}$ evaluated on \eqref{ansatz}-\eqref{5Dsol} vanish. Explicitly, the Lagrangian density on-shell reads
\begin{equation}
        16\pi G {\calL}_{\mathcal{M}} \equiv R - 2\Lambda + \alpha\left( R^2 - 4 R^{MN}R_{MN} + R^{MNPQ}R_{MNPQ} \right) =2 \left( \Lambda + \frac{3}{\ell^2} \right) + \frac{2\ttk}{L^2} \left( 1 - \frac{12\alpha}{\ell^2} \right) \nonumber
\end{equation}
while the integrand of the boundary term is
\begin{equation}
        8\pi G \mtsub{\calL}{\partial\calM} \equiv K + \alpha \left( J - 2 \hat{G}^{\mu\nu} K_{\mu\nu} \right) = \left( 1 + \frac{4\alpha\ttk}{L^2} \right) \frac{\sqrt{\nu^2+3}}{2\ell} \frac{(2r - r_+ - r_-)}{\sqrt{(r-r_+) (r-r_-)}} \,. \nonumber
\end{equation}
Both quantities vanish in virtue of (\ref{5Dsol}), to that $\mathcal{I}=0$. Something similar occurs in higher dimensions. This suggests that the thermodynamic properties of the WAdS$_3$ black holes in this theory are trivial, something we will confirm below by direct computation using different methods. This phenomenon is reminiscent of what happens with some Lifshitz black holes in higher-curvature gravity: In \cite{Cai:LifshitzBHR2} the authors found an asymptotically Lifshitz black hole with dynamical exponent $z=3/2$ in a 4-dimensional theory of gravity with both terms $R^2$ and $R^{\mu\nu}R_{\mu\nu}$ in the action. Such solution exhibits a vanishing on-shell action as well as vanishing entropy. The same happens for the WBH$_3\times \, \Sigma_{D-3}$ solution we construct here.

\subsection{Wald entropy formula}
To compute the entropy of the WBH$_3$ we resort to the Wald formula \cite{Wald:BHEntropy}, which amounts to compute the entropy by integrating a charge on the event horizon, $\mathcal{H}^+$. The formula for the entropy in $D$-dimensions reads
\begin{equation}
\label{WaldEntropy}
    \txsub{\calS}{W} = -2\pi \int_{\mathcal{H}^+} d^{D-2}x \sqrt{\sigma}\; \epsilon^{MN}\epsilon^{PQ} \frac{\partial \mathcal{L}}{\partial R^{MNPQ}} \,,
\end{equation}
where $\mathcal{L}=\mathcal{L}_{\mathcal{M}}+\mathcal{L}_{\partial \mathcal{M}}$ is the quadratic gravity Lagrangian, including boundary terms, $\epsilon_{MN}$ is the binormal tensor on $\mathcal{H}^+$, and $\sigma$ is the determinant of the induced metric on the constant-$t$ and constant-$r$ hypersurfaces evaluated on $\mathcal{H}^+$. Since the geometry is of the form WBH$_3\otimes \Sigma_{D-3}$, the integral in (\ref{WaldEntropy}) is over $D-2$ dimensions as it includes the angular direction $\varphi$ as well as the $D-3$ directions of the internal manifold $\Sigma_{D-3}$. 

Explicit computation of this charge integral yields vanishing entropy, $\txsub{\calS}{W} =0$. More explicitly, we get
\begin{equation}
     \epsilon^{MN}\epsilon^{PQ} \frac{\partial L}{\partial R^{MNPQ}} \propto  1 - 4\alpha\Lambda \ttk \,,
\end{equation}
which vanishes in virtue of \eqref{5Dsol}. That is to say, the entropy of the WBH$_3$ in this theory is identically zero.

\subsection{Noether-Wald conserved charges}

One can also compute the Noether charges of the solutions associated to translation invariance in $t$ and $\varphi $. To do that, one can consider the Iyer-Wald formalism \cite{Iyer.Wald:NoetherCharges, Iyer.Wald:NoetherEuclidean} which leads to an expression for the charges that takes the form
\begin{equation}
\label{NoetherWaldCharges}
    \txsub{\calQ}{W}[\xi] = \int_{S_\infty} d^{D-2}x \sqrt{\sigma}\: \epsilon_{MN} \, {Q}[\xi] ^{MN} \,,
\end{equation}
with ${Q}[\xi]$ being given by the Noether 2-form charge associated to the on-shell conserved current ${J}[\xi] = d{Q}[\xi]$; see \cite{Iyer.Wald:NoetherCharges} for details of its definition. $\xi$ is the Killing vector that generates the associated symmetry\footnote{The full expression for the Noether-Wald charge includes an extra term $-\xi \cdot {B}$ coming from the action boundary terms relevant for asymptotically flat spacetimes, cf. \cite{Iyer.Wald:NoetherEuclidean}. For asymptotically AdS spaces the B term cancels out when performing a background subtraction; see \cite{Dutta.Gopakumar:EntropiesAdS} for more details. Here we will make the same assumption but for asymptotically WAdS spaces.}. The integral is performed at a constant-$t$ and constant-$r$ hypersurface at infinity. 

As for asymptotically AdS spaces, the charges \eqref{NoetherWaldCharges} need to be regularized. One can do so by considering the background subtraction \cite{Dutta.Gopakumar:EntropiesAdS}. Following \cite{RiveraBetancour.Olea:Charges.CriticalGravity}, a more explicit expression for the Noether-Wald charge can be written down; namely
\begin{equation}
\label{NoetherWaldChargesQ}
    \txsub{\mathcal{Q}}{W}[\xi] = \int_{S_\infty} d^{D-2}x \sqrt{\sigma}\: \epsilon^{MN} \frac{\partial \mathcal{L}}{\partial R^{MNPQ}} \nabla^P\xi^Q \,,
\end{equation}
which, as for the Wald entropy formula, the integral goes over $\varphi $ and the coordinates on $\Sigma_{D-3}$. Applying this method to our solution WBH$_3 \otimes \mathbb{H}_2/\Gamma$, we get
\begin{equation}
M\equiv  \txsub{\mathcal{Q}}{W}[\partial_t ] = 0\, , \ \ \  J\equiv  \txsub{\mathcal{Q}}{W}[\partial_{\varphi } ]  = 0\, .
\end{equation}
More explicitly, for these Killing vectors we get
\begin{equation}
\label{EnergyWAdS}
	\epsilon^{MN} \frac{\partial L}{\partial R^{MNPQ}} \nabla^P \xi^Q \propto  1 - 4\alpha\Lambda \ttk \,,
\end{equation}
which vanishes in virtue of \eqref{5Dsol}. This means that, as for the entropy, both the mass and angular momentum of the WBH$_3$ of this theory are zero.


%
\subsection{Quasi-local stress-energy tensor}

Another method to compute the conserved charges is by means of the Brown-York quasi-local stress-energy tensor $\txsup{T}{(BY)}_{\mu\nu}$, cf. \cite{Brown.York:Stress.Tensor}. This method, frequently used in holographic renormalization in AdS, amounts to define the Brown-York stress-energy tensor near the boundary, adding counterterms to renormalize it, and then integrate its projection contracted with the Killing vector that defines an asymptotic isometry (see (\ref{BYcharge}) below). For the Einstein-Gauss-Bonnet theory, the renormalized boundary stress-energy tensor takes the form \cite{Davis:EGB.JunctionConditions, Brihaye.Radu:EGBblackholes}
\begin{equation}
    \label{renormTij}
    {T}_{\mu\nu} = \txsup{T}{(BY)}_{\mu\nu} + \txsup{T}{(ct)}_{\mu\nu} = 
    -\dfrac{2}{\sqrt{-h}} \dfrac{\delta {\calI} }{\delta h^{\mu\nu}} -\dfrac{2}{\sqrt{-h}} \dfrac{\delta \txsub{\calI}{ct}}{\delta h^{\mu\nu}} \,,
\end{equation}
where ${\calI}={\calI}_{\mathcal{M}}+{\calI}_{\partial\mathcal{M}}$ and so 
\begin{equation}
\label{EGBTBY}
    \txsup{T}{(BY)}_{\mu\nu} =\frac{1}{8\pi G} \left( K_{\mu\nu} - K h_{\mu\nu}  + 2 \alpha ( 3J_{\mu\nu} - J h_{\mu\nu} + 2 \hP_{\mu\rho\sigma\nu} K^{\rho\sigma} ) \right) \,,
\end{equation}
with
\begin{equation}
    \hP_{\mu\nu\rho\sigma} = \hR_{\mu\nu\rho\sigma} - 2 \hR_{\mu[\rho}h_{\sigma]\nu} + 2 \hR_{\nu[\rho}h_{\sigma]\mu}  + 2 \hR h_{\mu[\rho}h_{\sigma]\nu}  \,.
\end{equation}
$\txsub{\calI}{ct}$ are the counterterms in the action, which take the form 
\begin{equation}
\label{counterterms}
    \txsub{\calI}{ct} = \int_{\partial\calM} d^{D-1}x \sqrt{-h}\; \left( \alpha_0 + \alpha_1 \hat{R} + \alpha_2 \hat{R}^2 + \beta_{2} \hat{R}^{\rho\sigma} \hat{R}_{\rho\sigma} + \cdots \right) \,,
\end{equation}
with the hatted quantities referring to curvature tensors constructed with the boundary metric $h_{\mu\nu}$. The counterterms are necessary to regularize the infrared divergences due to the non-compactness of the spacetime. The conserved charges are defined as
\begin{equation}
\label{BYcharge}
	\calQ_{\text{BY}}[\xi] = \int_{S_\infty} d^{D-2}x \; \sqrt{\sigma} \; u^M \, {T}_{MN}\, \xi^N \,,
\end{equation}
where $\xi$ is a Killing vector and $u$ is the normal vector to the constant-$t$ codimension-2 surfaces $S_{\infty }$ at infinity, $r=\infty $. The integral in (\ref{BYcharge}) goes over $D-2$ dimensions, excluding time and the radial direction. The solution is a product WBH$_3\otimes \Sigma_{2}$, and so the computation of the gravitational energy by integrating on the angular coordinate $\varphi$ of the 3-dimensional spacewould actually give an energy-momentum density which is constantly extended along the directions of $\Sigma_2$. In fact, for the transverse directions $x^1$, $x^2$ we get 
\begin{equation}
\label{5DTBY}
	 T_{x^i}^{x^j} =   \dfrac{\sqrt{\nu^2+3}}{8\pi G\ell}\, \delta_{i}^{j}\,,
\end{equation}
with $i=1,2$. The integral of the regularized quasi-local stress-tensor over a codimension-2 spacelike surface of the the full space gives the total energy-momentum, which in this case vanishes: In fact, despite the non-zero components (\ref{5DTBY}), the relevant components in the integrand of the conserved charge (\ref{BYcharge}) associated to the symmetries generated by Killing vectors $\partial_t$, $\partial_{\varphi }$ are identically zero.

\section{Higher-dimensions and other compactifications}

Now, let us study the higher-dimensional case, which in particular allows for more general compactifications. We will consider different examples below.

\subsection{\texorpdfstring{WAdS$_3\times \mathbb{H}_{D-3}$}{WAdS3 x H(D-3)} vacua in \texorpdfstring{$D$}{D} dimensions}

Let us start by extending the 5-dimensional solution we studied above to $D$ dimensions by simply considering a $(D-3)$-dimensional internal space of constant curvature $\ttk$. One rapidly notices that the field equations demand $\ttk=-1$, so that $\Sigma_{D-3}$ ends up being locally equivalent to a $(D-3)$-dimensional hyperbolic space with metric (\ref{Sigma}) and curvature radius $L$. That is to say, the full space is of the form WBH$_3 \otimes \mathbb{H}_{D-3}/\Gamma$, i.e. locally WAdS$_3 \otimes \mathbb{H}_{D-3}$. Table \eqref{tab:table} summarizes the values for the parameters $ \alpha\,, \Lambda\,, L\,, \ell$, and the relations among them for the first five cases.

\label{app:WAdSSigmadim}
\begin{table}[ht]
    \centering
    \begin{tabular}{|c|c|c|c|c|} 
    \hline
    $D$ & $\alpha/L^2$ & $\Lambda\, L^2$ & $L^2/\ell^2$ & $\Lambda \, \alpha$ \\
    \hline\hline
    5 & 1/4 & -1 & 1/3 & -1/4 \\
    \hline
    6 & 1/12 & -3 & 1 & -1/4 \\
    \hline
    7 & 1/24 & -11/2 & 5/3 & -11/48 \\
    \hline
    8 & 1/40 & -17/2 & 7/3 & -17/80 \\
    \hline
    9 & 1/60 & -12 & 3 & -1/5 \\
    \hline
    \end{tabular}
    \caption{Relations between the parameters of the theory and the solution (locally) of the form WAdS$_3\times \mathbb{H}_{D-3}$ in $D$ dimensions. The curvature radius of WAdS$_3$ and $\mathbb{H}_{D-3}$ are $\ell$ and $L$ respectively.}
    \label{tab:table}
\end{table}

It is not difficult to obtain expressions for $\alpha$, $\Lambda$ and ${L^2}/{\ell^2}$ for arbitrary dimension $D$ (with $\ttk = -1$). These are given by
\begin{eqnarray}
\label{paramD}
    \dfrac{\alpha}{L^2} &=& \dfrac12 \frac{1}{(D-3)(D-4)} \\
    \Lambda L^2 &=& \frac{1}{4} \left(
    (D-5)(D-6) - 2(D-3)(D-4) \right) \\
     \frac{L^2}{\ell^2} &=& \frac{2D-9}{3} \,.
\end{eqnarray}
The first two relations make the Lagrangian to vanish while the third one makes the trace of the field equations to vanish. Let us notice we can obtain from \eqref{paramD} an expression for the warped critical points in Lovelock gravity in any dimension
\begin{equation}
    \Lambda \alpha = -\dfrac{1}{4} \left[ 1 - \dfrac{1}{2}\dfrac{(D-5)(D-6)}{(D-3)(D-4)} \right] \,.
\end{equation}

From the table and the equations above we notice that for $D=5$ and $D=6$ WAdS$_3\otimes \mathbb{H}_{D-3}$ is a solution at the same point of the parameter space. This is due to the fact that the density $R_{ABCD}R^{ABCD}-4R_{AB}R_{AB}+R^2$ identically vanishes for spaces of three dimensions or less, and therefore the terms in the field equations that are proportional to that combination do not contribute for geometries that are direct products of 3-spaces. In contrast, for spaces of four dimensions $4$ or more the integrand of the 4-dimensional Euler density does contribute with a non-vanishing constant.

\subsection{\texorpdfstring{WAdS$_3 \times \Sigma_3$}{WAdS3 x Sigma3} vacua in \texorpdfstring{$D=6$}{D=6} dimensions}

Next, let us consider solutions WBH$_3\otimes \Sigma_{D-3}$ whose internal manifold, $\Sigma_{D-3}$, is not necessarily locally equivalent to a maximally symmetric space. Let us focus in the case $D=6$ as an example. In that case, we may consider different cases, including products $\Sigma_{3}=\Sigma_{2}\times S^1$. We can also consider more abstruse deformations of the hyperbolic space $\Sigma_{3}=\mathbb{H}_3$, for instance by considering the 6-dimensional Kleinian space WAdS$_3 \otimes {\text{WAdS}}_3$. Remarkably, in the latter case the deformation parameters of both warped spaces, $\nu_{1,2}$, are independent and arbitrary, while their curvature radii have to be equal. Table \eqref{tab:table6} summarizes the relations among the parameters of some 6-dimensional solutions.
\begin{table}[h]
    \centering
    \begin{tabular}{|c|c|c|c|c|} 
    \hline
    $\Sigma_3$ & $\alpha/L^2$ & $\Lambda\, L^2$ & $L^2/\ell^2$ & $\Lambda \, \alpha$\\
    \hline\hline
    $ \mathbb{H}_3$ & 1/12 & -3 & 1 & -1/4 \\
    \hline
    WAdS$_3$ & 1/12 & -3 & 1 & -1/4 \\
    \hline
    $\mathbb{H}_2 \times {S}^1$ & 1/4 & -3 & 1/3 & -1/4 \\
    \hline
    \end{tabular}
    \caption{Relations between the parameters of the theory for solutions of the form WAdS$_3 \times \Sigma_3$ vacua in $D=6$ dimensions. The curvature radius of WAdS$_3$ and $\Sigma_3$ are $\ell$ and $L$ respectively.}
    \label{tab:table6}
\end{table}

In all these cases the on-shell Lagrangian vanishes.
 
\subsection{\texorpdfstring{WAdS$_3 \times \Sigma_{2} \times \tilde{\Sigma}_{2}$}{WAdS3 x Sigma2 x Sigma2} vacua in \texorpdfstring{$D=7$}{D=7} dimensions}

Now, consider a 7-dimensional case, which enables to consider solutions of the form $\Sigma_4=\Sigma_2\otimes \tilde{\Sigma}_2$. Some cases are summarizes in table \eqref{tab:table7}. 
\begin{table}[ht]
    \centering
    \begin{tabular}{|c|c|c|c|c|c|} 
    \hline
    $\Sigma_4$ & $\alpha/\ell^2$ & $\Lambda\, \ell^2$ & $L_1^2/\ell^2$ & $L_2^2/\ell^2$ & $\Lambda \, \alpha$\\
    \hline\hline
    $\mathbb{H}_2 \times \mathbb{H}_2$ & 1/24 & -9/2 & 1/3 & 1/3 & -3/16 \\
    \hline
    $\mathbb{H}_2 \times \mathbb{T}^2$ & 1/12 & -3 & 1/3 & $ \mathbb{R}$ & -1/4 \\
    \hline
    \end{tabular}
    \caption{Relations between the parameters of the theory for solutions of of the form WAdS$_3  \times \Sigma_2 \times \tilde{\Sigma}_{2}$ in $D=7$ dimensions. The curvature radius of WAdS$_3$, $\Sigma_2$ and $\tilde{\Sigma}_2$ are $\ell$, $L_1$ and $L_2$ respectively.}
    \label{tab:table7}
\end{table}

No $\Sigma_4=\mathbb{H}_2\times S^2$ compactification of this sort exists. The case $\Sigma_4= \mathbb{H}_4$ was considered in table \eqref{tab:table}. In all the cases the on-shell Lagrangian vanishes.

\subsection{Deformations of \texorpdfstring{WAdS$_3 \ltimes \Sigma_3$}{WAdS3 x Sigma3} warping products}

So far, we have only considered direct products of the form WBH$_3\otimes \Sigma_{D-3}$, and so we could ask whether the degeneracy in the parameter space we have observed in such cases is prerogative of the solutions that are direct product of simple spaces. In order to explore other type of geometries, we will consider here a warped product WAdS$_3 \ltimes \Sigma_3$ in 6 dimensions, and with a more general deformation of the internal hyperbolic space. Consider first the product space \eqref{ansatz} in $D=6$ with $\Sigma_3=\mathbb{H}_3$ being written in coordinates
\begin{equation}
    ds^2=g^{(\Sigma )}_{ij}dx^idx^j= \frac{L^2}{y^2}\left( dy^2 + 2 dx dz \right) \,,
\end{equation}
with $x^1=x$, $x^2=y$, $x^3=z$. Now, consider the following deformation
\begin{equation}
\label{h3def}
    ds^2 = \frac{L^2}{y^2}\left( dy^2 + 2 dx dz \right) + \dfrac{F(t, y, z)}{y^2} dz^2 \,,
\end{equation}
where $t$ is the time coordinate of the WAdS$_3$ piece of the 6-dimensional space, and $F(t, y, z)$ is a profile function to be determined by the field equations. Replacing the ansatz \eqref{h3def} in \eqref{EGB.eom}, the only restriction for the deformation profile $F(t,y,z)$ comes from the $x,z$ component of the field equations. It yields
\begin{equation}
    \frac{(\ell^2 - 12 \alpha)}{2\ell^2 L^2} \left( y\frac{\partial F}{\partial y} - y^2 \frac{\partial^2 F}{\partial y^2} \right) + \frac{3}{2} \frac{(\nu^2-1)(\ell^2 - 8\nu^2\alpha - 12\alpha^2)}{(\nu^2+3)\ell^2} \frac{\partial^2 F}{\partial t^2}= 0 \,.
\end{equation}
The rest of the components of the field equations impose the following restrictions among the parameters
\begin{equation}
	 \Lambda=-\frac{3}{\ell^2}\;,\; \; \alpha=\frac{\ell^2}{12}\;,\; \; L=\ell  \,.
\end{equation}
Therefore, for $\nu^2\neq 1$, and provided $\nu \neq 0$, we obtain
\begin{equation}
\frac{\nu^2}{(\nu^2+3)}\frac{\partial^2 F}{\partial t^2}=0\, ,
\end{equation}
and so the deformation profile must be a linear function of the warped time; namely
\begin{equation}
    F(t,y,z) = F_0(y,z) + F_1(y,z)\, t \,,
\end{equation}
with $F_{0}$ and $F_{1}$ being arbitrary functions of $y$ and $z$. These solutions are closely related to $pp$-waves in AdS (a.k.a. AdS-waves), which are a special class of Siklos spacetimes. In the case $\nu^2=1$, the function $F$ must satisfy 
\begin{equation}
     y\frac{\partial F}{\partial y} - y^2 \frac{\partial^2 F}{\partial y^2} = 0 \,,
\end{equation}
which is solved by the profile function 
\begin{equation}
    F(t,y,z) = G_0(t,z) + G_2(t,z)\, y^2 \,,
\end{equation}
with $F_{0}$ and $F_{1}$ being arbitrary functions of $t$ and $z$. These correspond to the massless modes of AdS$_3$ waves, cf. \cite{AyonBeato:0904}. 

This shows that the degeneracy of this special point of the parameter space of the higher-dimensional theory persists even when one considers a more general type of geometries, even with some warped products. There are, however, other warped solutions that are more restrictive in the $t$-dependence; for example, if one tries to look for solutions of the form \eqref{ansatz} with a time-dependent warping factor $f(t)$ in front of the metric $g_{ij}^{(\Sigma )}$, then the field equations impose $f=\text{const}$, and this is why we needed to consider more involved time-dependent warping products such as (\ref{h3def}) in order to find non-trivial solutions.

\section{Conclusions}

Summarizing, we have studied critical points of WAdS$_3$/WCFT$_2$ correspondence, which are given by higher-curvature gravity models on a specific curve of the parameter space. We have found solutions of the form WAdS$_3 \times \Sigma_{D-3}$ for $D\geq 5$ which allows for arbitrary warping factor $\nu $, i.e. with arbitrary squashing/stretching deformation of the WAdS$_3$ piece, generalizing what happens with the dynamical coefficient of the anisotropic scale invariant Schr\"odinger and Lifshitz spaces at the CS point of Lovelock gravity. In other words, in the sector we have studied, the theory behaves effectively almost as a topological theory, in the sense that the coefficient that controls the squashing/stretching deformation of independent pieces of the manifold are arbitrary. Besides, while the ratios of the radii of the different submanifolds do get fixed by the field equations, the total volume of the $D$-dimensional space is also arbitrary. This type of degeneracy, which is actually common in critical points of higher-curvature theories, describes a sort of zero-mode associated to scale invariance, while the arbitrariness of the value of $\nu $ makes the solution to be, so to speak, insensitive to the shape. This is also observed in warped compactifications. 

The fact of having found a critical point in a gravity theory with second order field equations is interesting on its own right. Higher-derivative theories typically give raise to extra massive excitations that, at the critical point, coalesce with a massless mode, leading ipso facto to the emergence of new low decaying mode in the bulk. The latter, from the dual perspective, comes to source states that render the CFT non-unitary – e.g. this is, for example, what happens with the so-called Log-gravity at the chiral point of TMG, and examples in Critical Gravity in higher dimensions can be constructed--.  This usually requires the prescription of strong boundary conditions which suffice to render the theory dynamically trivial. In our setup, being a higher-curvature theory of second order, this is different: The vanishing of the entropy and the conserved charges associated to the WAdS$_3$ black holes implies that the Virasoro central charge and the Kac-Moody level of the dual WCFT$_2$ are zero, and we take this as evidence that the latter theory is trivial. In many respects, these critical points of WAdS$_3 \times \Sigma_{D-3}$ vacua are the squashed/stretched analogs of the AdS$_D$ Chern-Simons point of Lovelock gravity.

\section*{Acknowledgements}

This work was supported by CONICET and ANPCyT through grants PIP1109-2017 and PICT-2019-00303.

\bibliographystyle{toine}
\bibliography{cites}{}

\end{document}